\documentclass[]{manu} 

\usepackage[left=0.875in,right=0.875in,top=1in,bottom=1.25in]{geometry}
\usepackage{graphicx}
\usepackage[flushleft]{threeparttable}
\usepackage{amsfonts}
\usepackage{amsmath}
\usepackage{multirow}
\usepackage{makecell}
\usepackage[colorlinks=true, allcolors=blue]{hyperref}
\usepackage{subcaption}
\usepackage{hhline}
\usepackage{makecell}
\usepackage{booktabs}
\usepackage[ruled,vlined]{algorithm2e}
\usepackage{bm}
\usepackage{setspace}

\graphicspath{ {.} }

\newcommand{\etal}{\textit{et al}. }
\begin{document}

\title{A novel adversarial learning strategy for medical image classification}

\author[a]{Zong Fan}
\author[a]{Xiaohui Zhang}
\author[b]{Jacob A. Gasienica}
\author[c]{Jennifer Potts}
\author[d]{Su Ruan}
\author[e]{Wade Thorstad}
\author[e]{Hiram Gay}
\author[f]{Pengfei Song}
\author[g]{Xiaowei Wang}
\author[a,e,h]{Hua Li}
\affil[a]{Department of Bioengineering, University of Illinois at Urbana-Champaign, IL, USA}
\affil[b]{Carle Illinois College of Medicine, Champaign, IL, USA}
\affil[c]{Joint Department of Biomedical Engineering, University of North Carolina, Chapel Hill, NC, USA and North Carolina State University, Raleigh, NC, USA}
\affil[d]{Laboratoire LITIS (EA 4108), Equipe Quantif, University of Rouen, Rouen, France}
\affil[e]{Department of Radiation Oncology, Washington University in St. Louis, MO, USA}
\affil[f]{Department of Electrical and Computer Engineering, University of Illinois at Urbana-Champaign, IL, USA}
\affil[g]{Department of Pharmacology and Bioeng., University of Illinois at Chicago, IL, USA}
\affil[h]{Cancer Center at Illinois, Urbana, IL, USA}

\authorinfo{Send correspondence to Hua Li. E-mail: li.hua@wustl.edu}

\maketitle

\begin{abstract}
Deep learning (DL) techniques have been extensively utilized for medical image classification.
Most DL-based classification networks are generally structured hierarchically and optimized through the minimization of a single loss function measured at the end of the networks. 
However, such a single loss design could potentially lead to optimization of one specific value of interest but fail to leverage informative features from intermediate layers that might benefit classification performance and reduce the risk of overfitting.
Recently, auxiliary convolutional neural networks (AuxCNNs) have been employed on top of traditional classification networks to facilitate the training of intermediate layers to improve classification performance and robustness.
In this study, we proposed an adversarial learning-based AuxCNN to support the training of deep neural networks for medical image classification.
Two main innovations were adopted in our AuxCNN classification framework.
First, the proposed AuxCNN architecture includes an image generator and an image discriminator for extracting more informative image features for medical image classification, 
motivated by the concept of generative adversarial network (GAN) and its impressive ability in approximating target data distribution. 
Second, a hybrid loss function is designed to guide the model training by incorporating different objectives of the classification network and AuxCNN to reduce overfitting.
Comprehensive experimental studies demonstrated the superior classification performance of the proposed model. 
The effect of the network-related factors on classification performance was investigated.
\end{abstract}

\section{Introduction}
\label{sec:intro}

Medical image classification plays an important role in clinical diagnosis by accurately recognizing medical images into different types. Numerous image classification techniques
ranging from conventional machine learning (ML) techniques~\cite{humeau2019texture,chandrashekar2014survey,wang2012machine} to deep learning (DL)-based methods ~\cite{Shen2017deeplearninginmed} have been proposed in recent decades. Traditional methods extract hand-crafted features such as shape and texture information from images for classification~\cite{Mall2019GLCMBF,Priyanka2020FeatureEA,sasikala2008wavelet,Zhou2015DetectionOP}.
Shape features such as perimeters and diameters describe the geometric characteristics of selected regions of interest (ROIs) in images~\cite{mingqiang2008survey}.
Texture features describe the perceived texture and spatial distribution of intensities of image ROIs~\cite{humeau2019texture}, such as gray level co-occurrence matrices (GLCM)~\cite{Haralick1973TexturalFF}, 
gray-level run-length matrices (GLRLM)~\cite{Tang1998TextureII}, etc. Texture features can be extracted using either statistical methods in the raw image domain,~\cite{Haralick1973TexturalFF, ojala2002multiresolution} 
or transform-based methods in the transformed image domain~\cite{Acharya2016ThyroidLC,Zhou2015DetectionOP}.

One problem of the aforementioned conventional methods is that redundancy may exist in the extracted features, which potentially increases the risk of overfitting and degrades the image classification performance~\cite{chandrashekar2014survey,kira1992feature}.
Therefore, feature selection procedures are usually employed to reduce feature redundancy and increase feature sparsity without losing essential classification-relevant information
~\cite{kira1992feature,Urbanowicz2018ReliefBasedFS}.
Features can be selected in either a supervised or unsupervised manner~\cite{Khaire2019StabilityOF}.
The unsupervised feature selection methods use proxy measures to score and select good features based on dataset characteristics before modeling~\cite{Urbanowicz2018ReliefBasedFS}, mainly including Relief algorithm~\cite{kira1992feature},
correlation-based feature selection (CFS) algorithm~\cite{hall2000correlation} and minimum redundancy maximum relevance (mRMR) algorithm~\cite{peng2005feature}.
Differently, supervised feature selection techniques use feature subset search strategy to iteratively select features that are relevant to classification performance evaluated by pre-defined classifiers such as support vector machine (SVM)~\cite{cortes1995support}.
The selected features can be used as inputs for downstream classification models. Several popular supervised feature selection methods include recursive feature elimination (RFE)~\cite{guyon2002gene} and heuristic feature search algorithm ~\cite{Khaire2019StabilityOF}.

Although conventional classification methods using hand-crafted features require relatively low computational resources while being able to achieve reasonable performance with limited training samples
~\cite{Mahony2019DeepLV}, the selected features may not be optimal for the image classification tasks~\cite{Alzubaidi2021ReviewOD}. 
Recently, deep learning techniques have been excessively employed in image classification applications
and shown powerful abilities to automatically extract deep discriminative features from images to improve task performance~\cite{Domingues2019UsingDL,Shen2017deeplearninginmed,xu2019current,zhang2022automated,huang2020dianet}. Instead of defining hand-crafted features for image classification using traditional methods,
deep neural networks (DNN) hierarchically learn features from input data samples to infer the corresponding label by use of supervised learning~\cite{Alzubaidi2021ReviewOD}. 
Popular DNNs for image classification tasks mainly use the convolutional neural network (CNN)-based network architectures such as VGGNet~\cite{Simonyan2015VeryDC}, 
Inception Network~\cite{Szegedy2015GoingDW},
and variations of residual neural network (ResNet)~\cite{He2016DeepRL}, to name a few. These CNN-based methods have achieved great successes in various medical imaging classification applications~\cite{Saric2019CNNbasedMF,spanhol2016breast,Han2017BreastCM,zhang2022automated}. 

However, DL-based methods can suffer from the overfitting problem caused by several data-related issues such as the limited amount of training data, data imbalance, and low inter-class reliability~\cite{Johnson2019SurveyOD}, which are common problems of medical image data.
Recently, generative adversarial network (GAN)~\cite{Goodfellow2014GenerativeAN}, which was originally proposed as an image synthesis technique, has been increasingly adopted in image classification to help address the overfitting problem and improve the classification performance \cite{zhu2018generative,zhan2017semisupervised}.
GANs are composed of two trainable networks with adversarial purposes, a generative network to synthesize an image similar to an actual image given a latent vector and a discriminative network that aims to discriminate between an actual and synthesized image.
During the adversarial training process,
the generator learns the desired data distribution and gains the powerful ability to automatically extract deep representative image features, and
the discriminator tries to differentiate between real and synthesized images~\cite{Domingues2019UsingDL,Shen2017deeplearninginmed}. 
GAN-based classification methods have been mainly designed in two different ways.
Some methods directly tune the discriminator of a trained GAN as an image classifier~\cite{Varghese2017GenerativeAN,zhan2017semisupervised,zhu2018generative},
while the other methods adopt the generator to synthesize images as a data augmentation strategy to reinforce the classification performance~\cite{Waheed2020CovidGANDA,bowles2018gan,Lu2019GenerativeAN, ma2020mri}.
Nevertheless, these two types of methods disentangle the training of a GAN and a classifier, that is, the classifier and the GAN are trained individually, which is cumbersome and the classification performance is affected by the quality of the trained GAN.

Moreover, most classification neural networks are structured hierarchically and optimized through the minimization of a single loss function which tunes the model to the desired performance based on a particular evaluation metric at the final output layer. 
Such single loss design might fail to leverage the features of the intermediate layers with abundant ROI semantic information that could benefit the overall network performance~\cite{Lin2017FeaturePN,Jain2017NonconvexOF,xu2015multi}.
To address this problem, auxiliary convolutional neural networks (AuxCNNs) have been proposed and received widespread attention in the field of image classification~\cite{he2021deeply}. 
By introducing an additional auxiliary network on top of the classification network, 
AuxCNNs provide auxiliary information from the extracted intermediate features as additional regularization to train the intermediate layers of the classification network,
therefore, improving the overall classification performance~\cite{Tian2020NetworkReg}.
Some AuxCNNs introduce supervision to train the intermediate layers through the final output layer~\cite{fang2020spatial,Zhao2020ExploringSF}, while others learn multi-scale image features from each hierarchical intermediate layer and integrate them to support the training of the classification network~\cite{sinha2020multi, GUAN2020multilabel,Schlemper2019AttentionGN}.

The effectiveness of the auxiliary networks and the representation power of GAN inspire our work to introduce GAN as an auxiliary network in the image classification model.
In this study, we proposed a novel adversarial learning-based AuxCNN to support the training of a DL-based classifier.
The AuxCNN is designed as a GAN architecture including an image reconstruction network (image generator) 
and a discriminator network, 
which serves as the additional regularization for the classifier and is trained with the classifier in an end-to-end manner. 
We hypothesize that instead of learning only classification-oriented information, the extracted features also represent the image generation-oriented information, which could potentially benefit the classification performance. 
In addition, a hybrid loss function is designed to guide the training of the classification network and the AuxCNN considering their distinct network architecture designs and learning objectives. 
Experimental results demonstrated the superior classification performance of the proposed method in terms of common evaluation metrics including precision, sensitivity, specificity, and F1-score.
The effects of various network-related factors on classification performance were investigated as well.

The remaining paper is organized as follows. 
Section~\ref{sec:method} introduces the proposed auxiliary learning-based classification method. 
Section~\ref{sec:imple} describes the datasets and implementation details of our method. 
The experimental results are shown in Section~\ref{sec:result}.
The discussion and conclusion are described in Section~\ref{sec:disc} and Section~\ref{sec:conclusion}, respectively. 

\section{Methods}
\label{sec:method}

\subsection{The proposed adversarial learning-based image classification framework}
\label{subsec:framework}

\begin{figure}[h]
\centering
\includegraphics[width=0.6\textwidth]{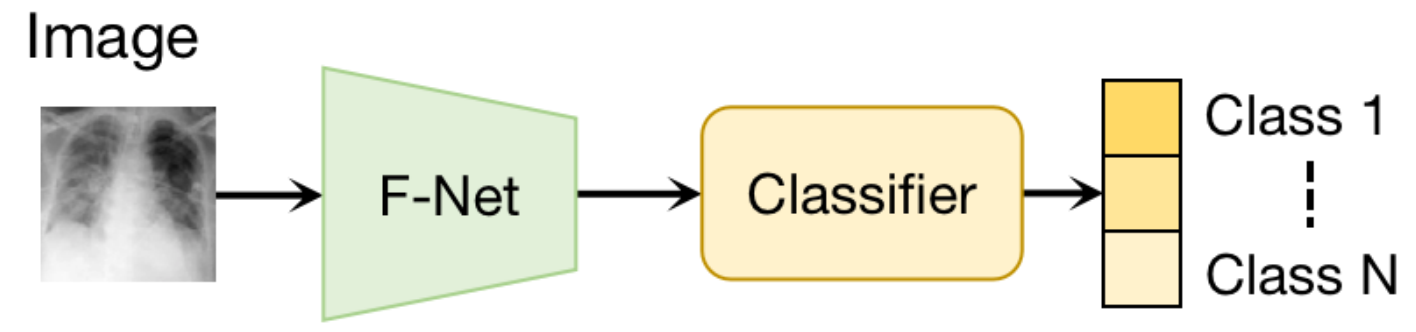}
\caption{The conventional image classification framework.} 
\label{fig:model_framework}
\end{figure}

The conventional DNN-based image classification models, as shown in Figure~\ref{fig:model_framework}, 
can be considered as the integration of a feature extraction network (F-Net) and a classifier.
Features of an input image are first extracted by F-Net, and used as the input of the classifier to estimate the probability of each class label.
Given a set of labeled training samples, F-Net and the classifier are optimized via an end-to-end training manner 
by minimizing a loss function, 
which specifies the penalty of deviations between the predicted outcomes and ground-truth class labels.
Gradient descent algorithms are the most commonly used optimization strategy.
However, this training process usually tries to solve a highly non-convex optimization problem.
The single loss design can potentially trap the optimization of the hierarchical network into the unsatisfying local minimum and jeopardize the overall network performance~\cite{Jain2017NonconvexOF,xu2015multi}.

\begin{figure}[h]
\centering
\includegraphics[width=0.7\textwidth]{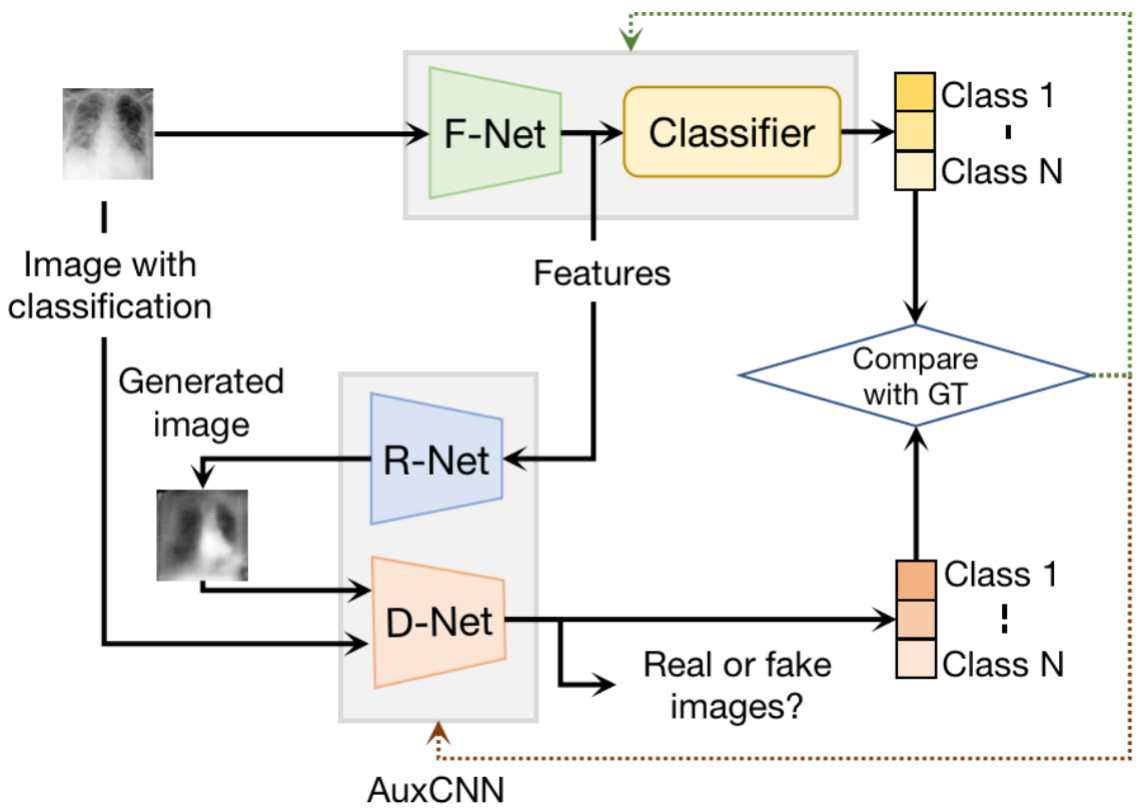}
\caption{The proposed adversarial learning framework. 
The green trapezoid represents the F-Net. 
The blue and orange trapezoids denote the two auxiliary networks of the proposed AuxCNN, the R-Net and D-Net, respectively. The solid arrows represent the forward operation flow; the dotted arrows represent the back-propagation flow.}
\label{fig:model_training}
\end{figure}

In this study, an AuxCNN-supported classification network is proposed to address these limitations.
As shown in Figure~\ref{fig:model_training}, 
the proposed framework includes an F-Net, a classifier, and an AuxCNN.
The F-Net and the classifier function the same as those in the conventional CNN-based classification networks.  
Motivated by the powerful ability of GANs to extract deep image features with rich distribution information, 
the AuxCNN is designed as the integration of a generator (R-Net) and a discriminator (D-Net) 
to support the training of F-Net and classifier.
The R-Net tries to reconstruct the input images using the features extracted by the F-Net, which allows the feature extractor to learn the distribution of input images.
The D-Net then distinguishes the reconstructed image from the original input images of the F-Net and classifies the image class label, ensuring the reconstructed images are semantically similar to the input images. 
In this way, the AuxCNN aims to enrich the extracted features of the F-Net with meaningful distribution-related information about the target image via the image reconstruction process to support the classification. 

Given a 2D image $X\in\mathbb{R}^{M\times M}$,
F-Net extracts a feature map $\mathbf{f}\in \mathbb{R}^{N_f}$ from $X$, 
and the classifier maps $\mathbf{f}$ to a vector $\mathbf{Y}=[Y_1, Y_2,...,Y_K]^T$, where $M\times M$ is the input image size, $N_f$ is the dimension of extracted feature, $K$ is the total number of classes, and 
each element $Y_k$ in $\mathbf{Y}$ represents the probability of $X$ belonging to the $k$th class. Feature map $\mathbf{f}$ and probability vector $P(\mathbf{Y}|X)$ can be represented as:
\begin{equation}
\begin{cases}
\mathbf{f} = F\left(X, \Theta_F\right) \\
P\left(\mathbf{Y}|X\right)=C\left(\mathbf{f},\Theta_{C}\right)
\end{cases},
\end{equation}
where $F$ represents the mapping function of F-Net with trainable parameters $\Theta_F$,
and $C$ is the mapping function of the classifier with trainable parameters $\Theta_C$.
The R-Net generates an estimation $\hat{X}$ of input image $X$ using the extracted feature $\mathbf{f}$:

\begin{equation}
\hat{X} = R\left(\mathbf{f},\Theta_R\right),
\end{equation}
where $R$ represents the mapping function of R-Net with trainable parameters $\Theta_R$.
D-Net is designed on top of R-Net.  
The input of D-Net is an image $X'$ that can be an original image or an image reconstructed via R-Net, that is, $X'\in \{X, \hat{X}\}$. 
D-Net contains a down-sampling network $D$ followed by two parallel network heads, the discrimination head $D_{disc}$ and classification head $D_{cls}$.
$D$ extracts a feature map $\mathbf{f}_D$ from $X'$.
$D_{disc}$ uses $\mathbf{f}_D$ to estimate discrimination probability $P(\mathbf{Y_{disc}}|X')$ and determine whether $X'$ is an original or reconstructed image. $D_{cls}$ uses $\mathbf{f}_D$ to estimate the classification probability $P(\mathbf{Y_{cls}}|X')$ and determine the class to which $X'$ belongs.
D-Net could be formulated as follows:

\begin{equation}
\begin{cases}
X'\in\{X,\hat{X}\}\\
\mathbf{f}_D = D\left(X',\Theta_{D}\right)\\
P\left(\mathbf{Y_{disc}}|X'\right)=D_{disc}\left(\mathbf{f}_D, \Theta_{disc}\right)\\
P\left(\mathbf{Y_{cls}}|X'\right)=D_{cls}\left(\mathbf{f}_D, \Theta_{cls}\right)
\end{cases},
\end{equation}
where $D$, $D_{disc}$ and $D_{cls}$ are the mapping functions of the feature extraction backbone of D-Net, the discrimination head and the classification head parameterized by trainable parameters $\Theta_D$, $\Theta_{disc}$, and $\Theta_{cls}$, respectively. 
After training, only the trained F-Net and the classifier are used to classify unseen images.

\subsection{Adversarial learning strategy}
\label{subsec:train}

An adversarial learning strategy is developed to train the framework shown in Figure~\ref{fig:model_training} by minimizing an integrated loss function $L_{cmb}$ expressed in Equation~(\ref{eq:cmb}):
\begin{equation}
\label{eq:cmb}
L_{cmb}\left(\Theta_F,\Theta_C,\Theta_R, \Theta_{D}, \Theta_{D_{disc}}, \Theta_{D_{cls}}\right) 
= L_{cls}\left(\Theta_F,\Theta_C\right) + \lambda_1 L_{rec}\left(\Theta_R\right) + \lambda_2 L_{adv}\left(\Theta_D, \Theta_{D_{disc}}, \Theta_{D_{cls}}\right),
\end{equation}
where $L_{cls}$, $L_{rec}$, and $L_{adv}$ represent the classification loss, reconstruction loss and adversarial loss, respectively; 
Each $\Theta$ represents the trainable parameters of each network as described above, 
and $\lambda_1,\lambda_2\in[0,1]$ are weighting factors which control the contribution of $L_{rec}$ and $L_{adv}$ to $L_{cmb}$, respectively.

Each loss function is designed specifically 
according to the different purposes of corresponding networks. 
The classification loss $L_{cls}$ employed for optimizing the F-Net and classifier is defined as a cross-entropy loss as shown in Equation~(\ref{eq:class-loss}):
\begin{equation}
    \label{eq:class-loss}
    L_{cls} = \mathbb{E}\left[-\mathbf{\bar{Y}}^T log\left(\mathbf{P}\right)\right],
\end{equation}
where $\mathbf{P}=[P(Y_1|X),P(Y_2|X),...,P(Y_K|X)]^T$ is the output vector of the classifier which represents the probabilities of the input image $X$ belonging to each of the classes, 
and $\mathbf{\bar{Y}}^T$ is the one-hot ground-truth (GT) label vector.  
The reconstruction loss $L_{rec}$ is defined in Equation~(\ref{eq:rec-loss}):
\begin{equation}
\label{eq:rec-loss}
\begin{aligned}
L_{rec} &= \left(1-\mathrm{SSIM}\left(X,\hat{X}\right)\right)/2 \\
            &= \left(1-\cfrac{\left(2\mu_X\mu_{\hat{X}}+\epsilon_1\right)\left(2\sigma_{X\hat{X}}+\epsilon_2\right)}{\left(\mu_X^2+\mu_{\hat{X}}^2+\epsilon_1\right)\left(\sigma_X^2+\sigma_{\hat{X}}^2+\epsilon_2\right)}\right)/2,
\end{aligned}
\end{equation}
where $\mathrm{SSIM}$ is the structural similarity index (SSIM) between the input image $X$ and the reconstructed image $\hat{X}$~\cite{Wang2004ImageQA}, which could benefit the training of discriminator by relaxing the constraint in finding the optimal parameters~\cite{Su2019OGANEC}.
$\mu_X$ and $\sigma_{X}$ denote the mean and variance of input image $X$. 
$\mu_{\hat{X}}$ and $\sigma_{\hat{X}}$ are the mean and variance of reconstructed image $\hat{X}$, and
$\sigma_{\mathit{X\hat{X}}}$ is the covariance of $X$ and $\hat{X}$.
Here $\epsilon_1$ and $\epsilon_2$ are small constants to stabilize the division operation.
$L_{adv}$ is defined as the integration of two losses $L_{adv}^{disc}$ and $L_{adv}^{cls}$, 
where $L_{adv}^{disc}$ measures the output of the discrimination head of D-Net, 
and ${L_{adv}^{cls}}$ measures the output of the classification head of D-Net. 
Cross-entropy is employed for both $L_{adv}^{disc}$ and $L_{adv}^{cls}$, as shown in Equation~(\ref{eq:adv-loss}):

\begin{equation}\label{eq:adv-loss}
\begin{aligned}
L_{adv} &= \lambda L_{adv}^{disc} + \left(1-\lambda\right) L_{adv}^{cls};\\
L_{adv}^{disc} & =\mathbb{E}\left[logD\left(X\right)\right] + \mathbb{E}\left[log\left(1-D\left(R\left(F\left(X\right)\right)\right)\right)\right]; \\
L_{adv}^{cls} &= \mathbb{E}\left[-\mathbf{\bar{Y}}^T log\left(\mathbf{P'}\right)\right];
\end{aligned}
\end{equation}
where $\lambda$ represents the weighting factor to combine the two losses and $\mathbf{P'}=[P(Y_{cls}^{(1)}|X'),...,P(Y_{cls}^{(K)}|X')]^T$ is the output vector of classifier $D_{cls}$. 
The detailed training procedure is described in Algorithm~\ref{algo:algorithm}. 
The trained F-Net and the classifier could be disentangled to perform forward propagation individually for classification. 
The predicted label of an input data $X$ is determined as the class label with the highest probability, which equals to:

$$Y=\underset{i}{argmax} P\left(Y_i|X\right), i\in\{1,2,..,k\}$$

\begin{algorithm}[!ht]
\SetAlgoVlined
\DontPrintSemicolon
\SetKw{param}{Networks with Paramters:}
\SetKw{hyper}{Training Hyperparameters:}
\SetKw{start}{Start Training}
\SetKw{stop}{End Training}
\KwIn{Training dataset $\mathcal{X}$ with total $N$ pairs of images and related GT class labels}
\param{\textnormal{F-Net with $\Theta_F$; classifier with $\Theta_C$; R-Net with $\Theta_R$; D-Net with $\Theta_D, \Theta_{disc}, and\ \Theta_{cls}$}}

\hyper{\textnormal{Minibatch size: $m$; the number of training iterations: $t$}}

\vspace{0.25cm}
\start

Current training iteration: $i=1$\;
\While{$i \le t$}{
    The number of trained images in current epoch: $n$; Set $n=0$\;
    \While{$n\le N$}{
        \begin{minipage}{0.85\linewidth}
        \begin{enumerate}
        \item Sample minibatch of $m$ images $X=\{X^{(1)},X^{(2)},...,X^{(m)}\}$ and their corresponding class labels $\bar{Y}=\{\bar{Y}^{(1)},\bar{Y}^{(2)},...,\bar{Y}^{(m)}\}$ from dataset $\mathcal{X}$. Set the original or reconstructed discrimination labels of $X$ as $\bar{Y}_{disc}=\{1, 1, ..., 1\}$ with number of $m$ labels\;
        \item Forward $X$ through the F-Net and R-Net to obtain reconstructed images $\hat{X}$. Set the original or reconstructed discrimination labels as $\hat{Y}_{disc}=\{0, 0, ..., 0\}$\;
        \item Combine both original and reconstructed images $X'\in \{X, \hat{X}\}$ and forward $X'$ through D-Net to get outputs of classification prediction $Y_{cls}'$ and discrimination prediction $Y_{disc}'$\;
        \item Calculate the $L_{adv}$ as Equation~(\ref{eq:rec-loss}) by comparing $Y_{cls}'$ and $Y_{disc}'$ to $\bar{Y}$ and $\bar{Y}_{disc}$, respectively\;
        \item Calculate the partial derivatives of each parameter of D-Net: $\bigtriangledown_{\Theta_D, \Theta_{disc}, \Theta_{cls}} L_{adv}$\;
        \item Update the $\Theta_D, \Theta_{disc}, \Theta_{cls}$ of D-Net by ascending the gradients calculated in previous step, while fixing the parameters of F-Net, classifier, and R-Net\;
        \item Repeat step 2 and 3\; 
        \item Forward $X$ through F-Net and classifier to get classification prediction $Y$\;
        \item Calculate the combined loss $L_{cmb}$ as Equation~(\ref{eq:cmb})\;
        \item Calculate the partial derivatives of parameters of the F-Net, classifier, and R-Net: $\bigtriangledown_{\Theta_F, \Theta_C, \Theta_G}L_{cmb}$\;
        \item Update the F-Net, classifier and R-Net by ascending the gradients while fixing the parameters of the D-Net\;
        \item n=n+m\;
        \end{enumerate}
        \end{minipage}
    }
    Iteration $i=i+1$\;
}
\stop

\KwOut{F-Net and classifier with trained parameters $\Theta_F$ and $\Theta_C$}
\caption{Minibatch training of the proposed training framework}
\label{algo:algorithm}
\end{algorithm}

\section{Dateset \& Method Implementation}
\label{sec:imple}

\subsection{Dataset}
\label{subsec:dataset}
The proposed framework was evaluated using two different datasets, a public COVID-19 patient dataset called COVIDx~\cite{wang2020covid} and an oropharyngeal squamous cell carcinoma (OPSCC) patient dataset~\cite{malia2022learning}.

\subsubsection{COVID-19 dataset}
\begin{figure}[h]
\centering
\includegraphics[width=0.5\textwidth]{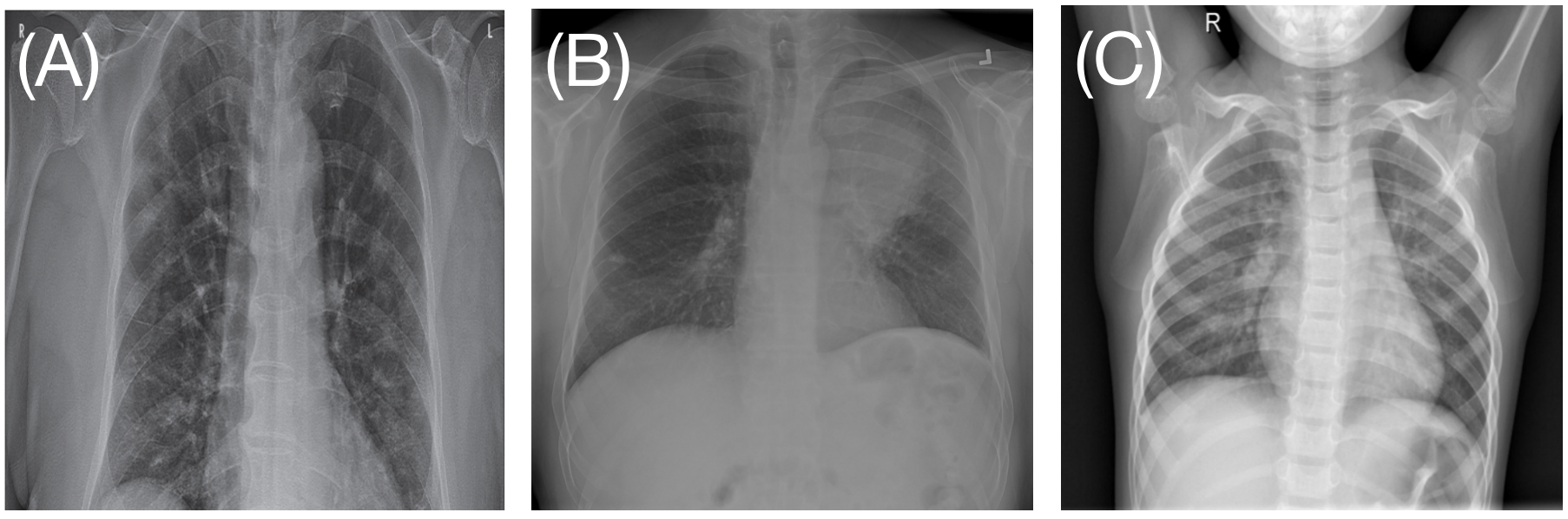}
\caption{Example CXR images of (A) COVID-19 positive case, (B) non-COVID-19 pneumonia case and (C) normal healthy case in the COVIDx dataset.}
\label{fig:COVIDx_image}
\vspace{0.1in}
\end{figure}

The public COVID-19 patient dataset, COVIDx, was used to evaluate the performance for the task of multi-class classifications.
This dataset includes 13,958 2D chest X-ray (CXR) images collected from 13,870 patients,
with the diagnosis of three classes of COVID positive, non-COVID pneumonia, and normal cases. 
Example images of each diagnosis are shown in Figure~\ref{fig:COVIDx_image}.
The raw image size is $1024\times 1024$ pixels. 
Details of this dataset are shown in Table~\ref{tab:covidx_dataset}.
This dataset shares the common data-related issues in medical image classification tasks.
Only a few hundred images are determined as COVID-19 positive. 
Severe data imbalance exists where the number of COVID-19 images is much smaller than that of non-COVID pneumonia images. 
In addition, COVID-19 positive cases and non-COVID pneumonia cases have very high visual similarities, 
which makes an accurate classification of COVID-19 positive cases a challenging task. 
This dataset is suitable for evaluating the performance of our proposed strategy
on addressing the challenges described above.

\subsubsection{OPSCC dataset}
\begin{figure}[h]
\centering
\includegraphics[width=0.7\textwidth]{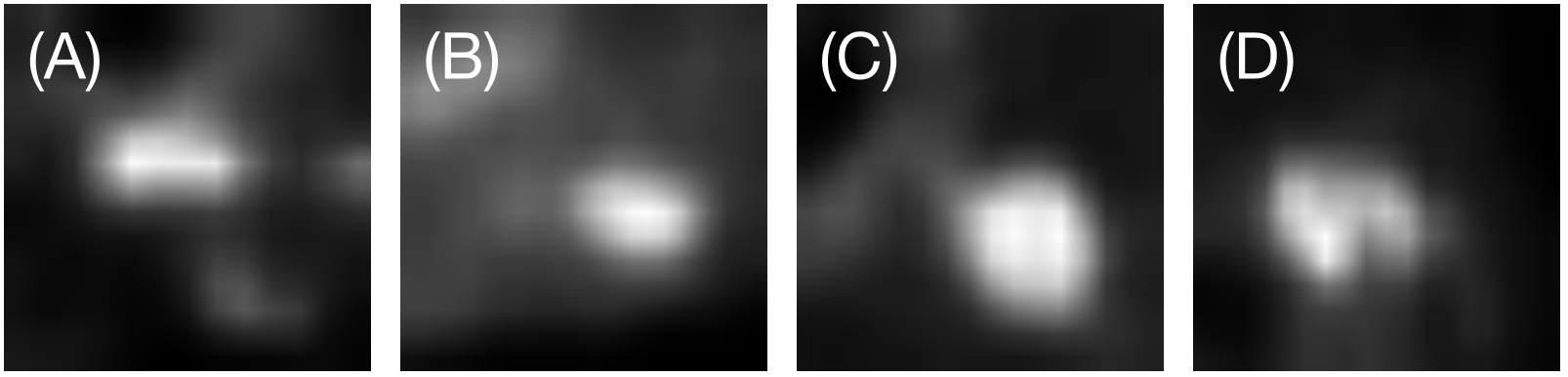}
\captionof{figure}{Example PET images of OPSCC dataset. (A) and (B) are from the patients with low-risks of treatment failure. (C) and (D) are from patients with high-risks of treatment failure.}
\label{fig:HN_image}
\end{figure}

The second dataset is a real-patient dataset collected in our previous study~\cite{malia2022learning}, which includes 3,255 2D positron emission tomography (PET) images acquired from 150 oropharyngeal squamous cell carcinoma (OPSCC) patients with low risks of treatment failure and 67 patients with high risks of treatment failure. Details of this dataset are shown in Table~\ref{tab:hn_dataset}.
The collection procedure follows the IRB protocol approved by the Human Research Protection Office of the Washington University School of Medicine in St. Louis. 
Several sample images are shown in Figure~\ref{fig:HN_image}
This dataset with low image quality could help test the potential of the proposed method for real clinical applications.

\begin{table}
\centering
\begin{threeparttable}
\captionof{table}{Details of the COVIDx dataset.\label{tab:covidx_dataset}}
\begin{tabular}{c c c}
\hhline{===}
\textbf{Class label} & \textbf{Training images} (\%) & \textbf{Testing images} (\%) \\
\hline
COVID-19 & 417 (3.1) & 100 (33.3)  \\
Normal & 7866 (57.6) & 100 (33.3)\\
Pneumonia & 5375 (39.3) & 100 (33.3) \\
\hline
Total & 13658 & 300 \\
\hhline{===}
\end{tabular}

\end{threeparttable}
\vspace{0.1in}
\end{table}

\begin{table}
    \centering
    \captionof{table}{Details of the OPSCC dataset.\label{tab:hn_dataset}}
    \begin{tabular}{c c c}
    \hhline{===}
    \textbf{Class label} & \textbf{Training images} (\%) & \textbf{Testing images} (\%) \\
    \hline
    High-risk & 905 (29.6) & 100 (50) \\
    Low-risk & 2150 (70.4) & 100 (50) \\
    \hline 
    Total & 3055 & 200\\
    \hhline{===}
    \end{tabular}
    \vspace{0.1in}
\end{table}

\subsection{Network architectures}
\label{subsec:architectures}

\begin{figure}[h]
\centering
\includegraphics[width=0.7\textwidth]{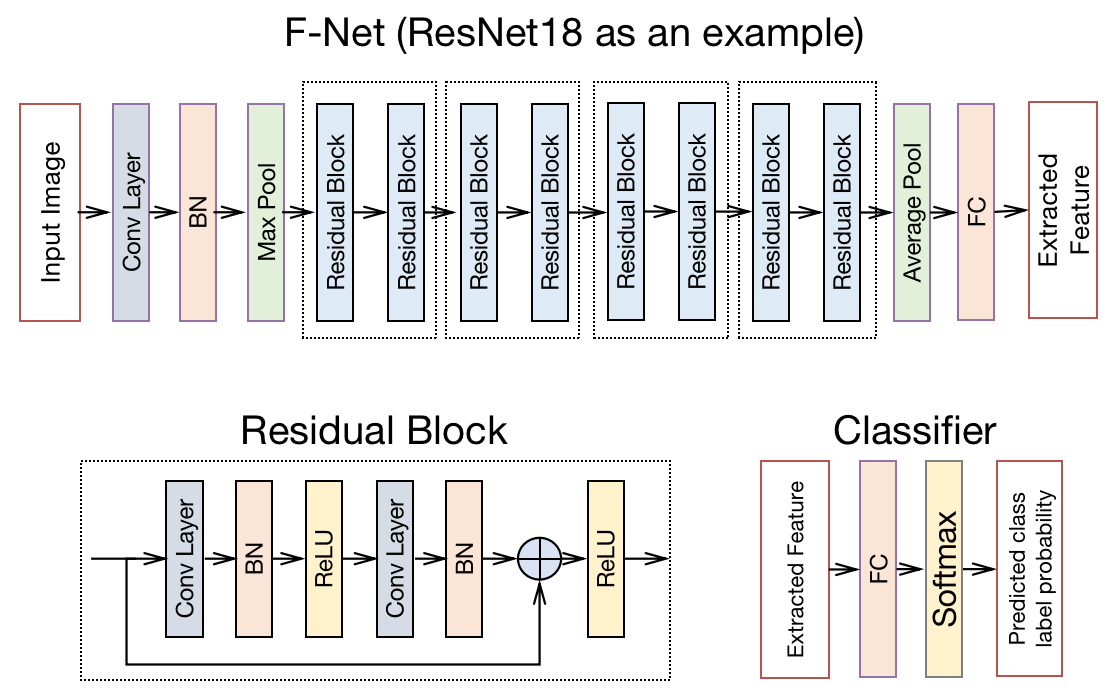}
\vspace{0.3in}
\caption{The designed F-Net architecture based on ResNet18. \textbf{BN} represents the batch normalization layer.}
\label{fig:F_Net_Architecture}
\end{figure}

Considering the model generalizability, F-Net is designed as a standard residual neural network (ResNet)~\cite{He2016DeepRL}, 
which is a commonly used network architecture in various medical image classification tasks~\cite{chen2017low,Saric2019CNNbasedMF,Domingues2019UsingDL,lopez2020transfer}.
Compared to vanilla CNNs, ResNets apply shortcut connections between non-adjacent convolutional (Conv) layers in residual blocks,
to address the gradient vanishing issue for deeper networks with more activation layers~\cite{He2016IdentityMI}. 
The architecture of ResNet was explained by the use of ResNet18 as an example shown in Figure~\ref{fig:F_Net_Architecture}.
Particularly, the final fully-connected (FC) layer of the original ResNet with $N_f$ neurons outputs the intermediate feature, which is used for either classification via the classifier or image reconstruction via the R-Net. 
The following classifier consists of a single FC layer with a number of $K$ neurons followed by a softmax activation layer, where $K$ is the number of class labels.

\begin{figure}[h]
\centering
\includegraphics[width=0.5\textwidth]{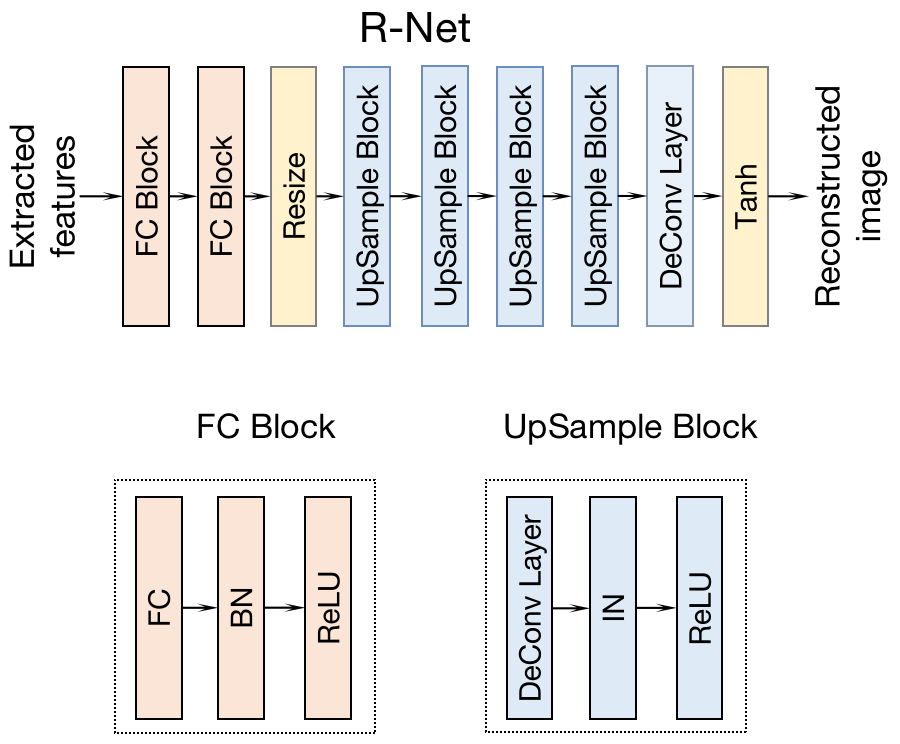}
\vspace{0.2in}
\caption{The network architecture of R-Net. \textbf{IN} represents the instance normalization layer.
}
\label{fig:R_Net_Architecture}
\end{figure}

\begin{figure}[h]
\centering
\includegraphics[width=0.5\textwidth]{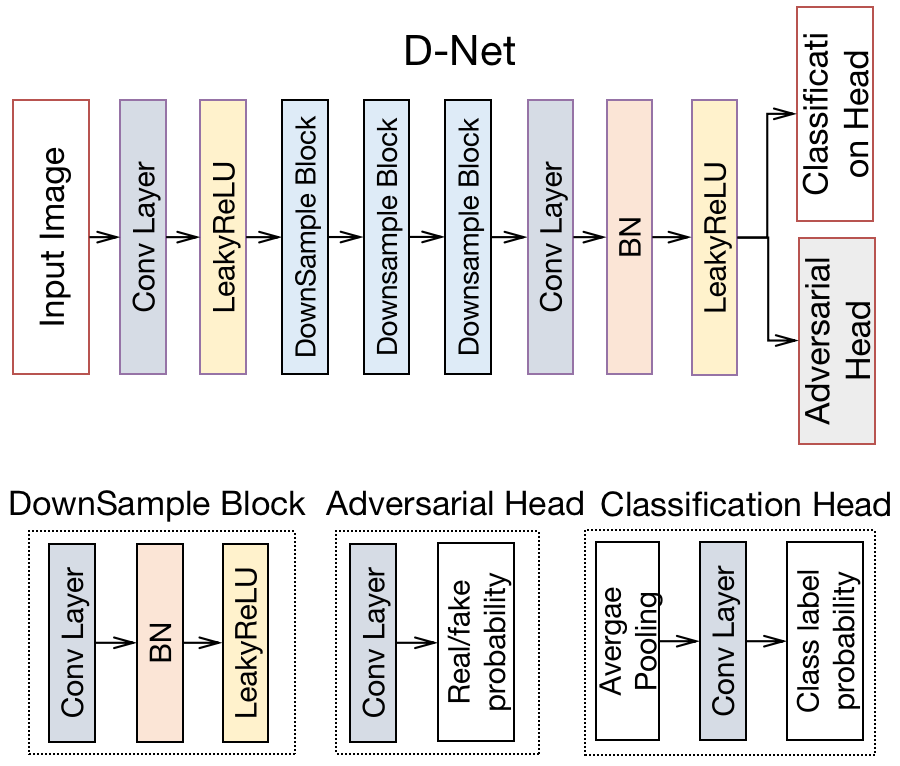}
\vspace{0.2in}
\caption{The network architecture of D-Net.
}
\label{fig:D_Net_Architecture} 
\end{figure}

The architecture of R-Net is adapted from InfoGAN,~\cite{Chen2016InfoGANIR}
which is one of the popular generative networks with simple structures but impressive conditional image generation capability.
As shown in Figure~\ref{fig:R_Net_Architecture}, 
the first two blocks of R-Net are FC blocks, where each block contains an FC layer followed by a batch normalization (BN) layer and a rectified linear activation (ReLU) layer. 
The third block is a reshape layer, which transforms the 1D vector generated by the adjacent FC block into a 3D feature map with the size of $C_r\times D_r\times D_r$, where $C_r$ represents the number of channels and $D_r$ is the width or height of the intermediate feature map. 
A set of $N_{up}$ UpSample blocks up-sample the 3D feature map by increasing the size of $2^{N_{up}}$ times. Each UpSample block contains a deconvolutional (DeConv) layer (kernel size of 4 and stride of 2), an instance normalization layer (IN), and a ReLU layer. 
The number of UpSample blocks is determined by the input image size, that is, $N_{up}=Log_2( M/ D_r)-1$, where $M$ is the input image size. 
Finally, a DeConv layer (kernel size of 4 and stride of 2) and a \textit{tanh} activation layer are employed on top of the last UpSample block 
to reconstruct an image with identical sizes as the input image of F-Net.

The architecture of discriminator $D$ is adapted based on the discriminative network in PatchGAN~\cite{isola2017image},
as shown in Figure~\ref{fig:D_Net_Architecture}. PatchGAN discriminator is a type of discriminator that tries to classify whether each local image patch is real or fake, which has good modeling capacity and computational efficiency. 
The first two layers consist of a Conv layer (kernel size of 4 and stride of 2) and a Leaky-ReLU activation layer.
The Leaky-ReLU layer sets a small slope for the negative input values from the adjacent Conv layer, rather than a flat slope in the regular ReLU layer~\cite{Xu2015EmpiricalEO}.
This activation layer is more suitable for training the adversarial network to deal with the sparse gradients problem.
Next, $N_{down}$ DownSample blocks are employed to down-sample the feature map progressively to reduce the feature map size by $2^{N_{down}}$ times.
Each DownSample block contains a Conv layer (kernel size of 4 and stride of 2), a BN layer, and a Leaky-ReLU layer. 
Finally, a Conv layer (kernel size of 3 and stride of 1), a BN layer, and a Leaky-ReLU layer are appended on top of the DownSample blocks to produce the intermediate feature $\mathbf{f}_D$ used for further classification and original/reconstructed discrimination. 
The discrimination head is comprised of a Conv layer and a sigmoid activation layer that produces a 2D feature map with the size of $M/2^{N_{down}+1}\times M/2^{N_{down}+1}$, where each pixel represents the probability of the corresponding local image patch of the input image being original or reconstructed. 
The classification head contains an average pooling layer, a Conv layer (kernel size of 1) and a softmax activation layer, which outputs a 1D vector with $K$ neurons. 
This functions similarly to the classifier to output the probability of the class label of the input image.
The use of the Conv layer instead of the FC layer in the two heads could help reduce the number of trainable parameters, thereby potentially reducing the computational resource cost and improving training efficiency. 

\subsection{Network parameter settings}
\label{subsec:net_hyper}

In the case study of three-class classification on the COVIDx dataset, the input image sizes $M\times M$ were set to three different settings of $M\in\{112,\ 224,\ 448\}$ to investigate the impact of image size on the classification performance of the proposed framework.
A family of ResNet architectures, including ResNet12, ResNet18, ResNet34, ResNet44, and ResNet50, were used as F-Net to evaluate the effect of network depths on the classification performance of the proposed method. 
For example, ResNet18 includes 8 residual blocks and 
each residual block contains 2 Conv layers interconnected with BN and ReLU layers, as shown in Figure~\ref{fig:F_Net_Architecture}. 
The number of the residual blocks when using other ResNets followed the design in the original ResNet paper~\cite{He2016DeepRL}. 
The last FC layer of the F-Net had $N_f=128$ neurons and the classifier had $K=3$ neurons in the FC layer.

For R-Net (shown in Figure~\ref{fig:R_Net_Architecture}), two FC layers in the first two FC blocks contained 1024 and $128\times D_r\times D_r$ neurons, respectively, where $D_r=7$. 
The following reshape layer transformed the 1D latent representation into a 3D feature map with shape as $128\times D_r\times D_r$. 
The number of UpSample blocks was $N_{up}$ was set to 3, 4, and 5 corresponding to the input image size $M\in\{112,\ 224,\ 448\}$, respectively.
Each UpSample block transformed the input feature to expand the size of the feature map by a factor of 2 and reduce the number of channels by a factor of 0.5. 

For D-Net, as shown in Figure~\ref{fig:D_Net_Architecture}, the number of DownSampling blocks was set to $N_{down}=3$. 
The number of output channels of the first Conv layer was 64 and was doubled after each DownSample block.
In the discrimination head, the number of output channels of the Conv layer was 1, while the number was 3 in the classification head. 

As for the binary classification on the OPSCC dataset, the F-Net was defined as the ResNet12 considering the small input image size. The input image size was set to $M=32$, and the classifier had $K=2$ neurons in the FC layer. 
The rest setup of the R-Net and D-Net were the same as those used in the COVID-19 classification study described above. 

\subsection{Parameter settings for framework training and testing}
\label{subsec:training}

The details of the framework training process are shown in Algorithm~\ref{algo:algorithm}.
To prepare the training and testing data, a set of images containing 100 examples from each class label was randomly selected from the whole dataset as the testing dataset. 
The rest of the images were randomly divided into the training dataset (80\%) and the validation dataset (20\%).
This separation method was employed in all experiments in this study.

In the COVD-19 case study, when F-Net was set as ResNet18, ResNet34, or ResNet50, 
the parameters were initialized using the parameters of the networks that were pre-trained on the ImageNet dataset~\cite{Deng2009ImageNetAL} and released in PyTorch community~\cite{paszke2019pytorch}.
For F-Net set as ResNet12 or ResNet44, the model parameters were initialized using the Xavier initialization scheme~\cite{pmlr-v9-glorot10a}. 
The initial parameters of R-Net and D-Net were determined using the same strategy.
In the OPSCC case study, all F-Nets were initialized following the Xavier initialization scheme.

In each training iteration, a mini-batch of 8 images ($m=8$) were sampled randomly from the training dataset. 
The images were then preprocessed by normalizing the pixel values to $[0, 1]$, followed by a data augmentation process, including random rotation of angle $\rho\in[-10^\circ,10^\circ]$ and random horizontal flip. 
The preprocessed images were subsequently resized to $M\times M$ pixels through bi-linear interpolation, where $ M\in\{112, 224, 448\}$ in the COVID-19 case study and $M=32$ in the OPSCC case study. 
The processed images were the input images of F-Net and D-Net in the proposed training framework.

To update the network parameters, the loss functions defined in Section~\ref{subsec:train} were minimized. 
The weighting factors $\lambda_1$ and $\lambda_2$ in Equation~(\ref{eq:cmb}) used for addressing the contribution of the reconstruction loss and adversarial loss were set to $0.2$ and $1.0$.
In the reconstruction loss, the small constant $\epsilon_1$ and $\epsilon_2$ in Equation~(\ref{eq:rec-loss}) were set to $10^{-6}$.
In the adversarial loss, the weighting factor $\lambda$ in Equation~(\ref{eq:adv-loss}) used for balancing the discrimination loss and classification loss was set to $0.5$.
Adam optimizer~\cite{Kingma2015AdamAM} was employed with decay rate $\beta_1=0.5,\beta_2=0.999$ and initial learning rate $lr=0.001$. 
The framework was trained on the training dataset for 200 epochs.

The proposed framework was implemented using PyTorch 1.7.0~\cite{paszke2019pytorch} and was trained and validated on Nvidia GeForce GTX 1080 Ti GPUs. 
After training, the model with the highest accuracy on the validation data was further tested on the testing dataset. 
To assess the network stability, the above procedure was repeated three times, from data selection to model testing. 
The mean and standard deviation of the results were calculated to evaluate the network performance.

\subsection{Implementation of other methods for comparison}
\label{subsec:comp}

Several other popular methods, 
including the methods of random over-sampling (ROS)~\cite{kubat1997addressing}, 
focal loss~\cite{Lin2017FocalLF}, 
and hard example mining (HEM)~\cite{Shrivastava2016TrainingRO}, have been implemented for comparison.
These methods have been proved to be effective in solving the overfitting problem due to the limited and imbalanced training dataset.
The ROS involves randomly selecting samples from the minority class and adding them to the training dataset, 
thus producing mini-batches with a balanced class distribution. 
This method is simple to implement while being effective in alleviating the data imbalance issue. 

Different from the ROS, the focal loss method optimizes the network with a focal loss instead of the traditional cross-entropy loss. 
Focal loss adjusts the penalty of hard examples in classification loss and highlights the contributions from the minority classes, where the hard examples are those images that are obviously mislabeled by the classifier.
It can alleviate the issues of dataset imbalance, especially when the images of different classes have high visual similarities. 
Focal loss is defined as~\cite{Lin2017FocalLF}:
\begin{equation}
\label{eq:fl}
\begin{aligned}
\mathrm{FL}\left(p_t\right) & = -\alpha_t\left(1-p_t\right)^\gamma log\left(p_t\right),\\
p_t = &
\begin{cases}
p & \text{if y=1}\\
1-p & \text{otherwise}\\
\end{cases},\\
\alpha_t = &
\begin{cases}
\alpha & \text{if y=1}\\
1-\alpha & \text{otherwise}\\
\end{cases},
\end{aligned}
\end{equation} 
where $y=1$ specifies the ground-truth label of an input image and $p\in[0,1]$ is the model's estimated probability of classification. 
The focusing parameter $\gamma$ adjusts the rate of how the easily-classified majority examples should be down-weighted. 
Larger $\gamma$ means a smaller contribution from the majority of examples when tuning the model parameters. 
The parameter $\alpha$ is the weighting factor to address the mislabeled examples. 
In this study, $\gamma$ was set to 1.5 and $\alpha$ was set to $0.25$ based on experimental trials.

HEM method aims to improve the performance of the trained network by augmenting the training dataset with data examples that have lower predicted classification accuracy.
In this way, the learned feature can avoid overfitting and be less biased by learning more from the minority and hard samples during training. 
In this study, the top 25\% hard examples with the highest classification errors in the training mini-batch were treated as the hard examples. 
These hard examples were used to augment the training dataset.

For the experiments shown below in Section~\ref{sec:result}, the network architectures of these three methods use the same F-Net and classifier described in Section~\ref{subsec:architectures}. 
The dataset, network, and training parameter settings are similar as discussed in Section~\ref{subsec:dataset} to Section~\ref{subsec:training}. 
Each of the three methods was adopted in training independently.

\subsection{Performance evaluation metrics}
\label{subsec:metric}

Precision (positive predicted value, PPV), specificity (true negative rate, TNR), sensitivity (true positive rate, TPR), F1-score (or dice coefficient) were employed as the metrics to evaluate the classification performance in all the case studies.
High sensitivity indicates a low probability of misdiagnosis if the prediction result is negative.  
High specificity indicates high confidence of infection when the prediction result is positive.
F1-score was the harmonic mean of precision and recall, representing the overall performance of a model. 
In addition, for the experiments using OPSCC dataset for binary classification, the area under the ROC curve (AUC) was also calculated.  

\section{Experimental Results}
\label{sec:result}

\subsection{Classification performance of the proposed framework}
\label{subsec:perf}

\begin{figure}[h]
    \centering
    \includegraphics[width=\textwidth]{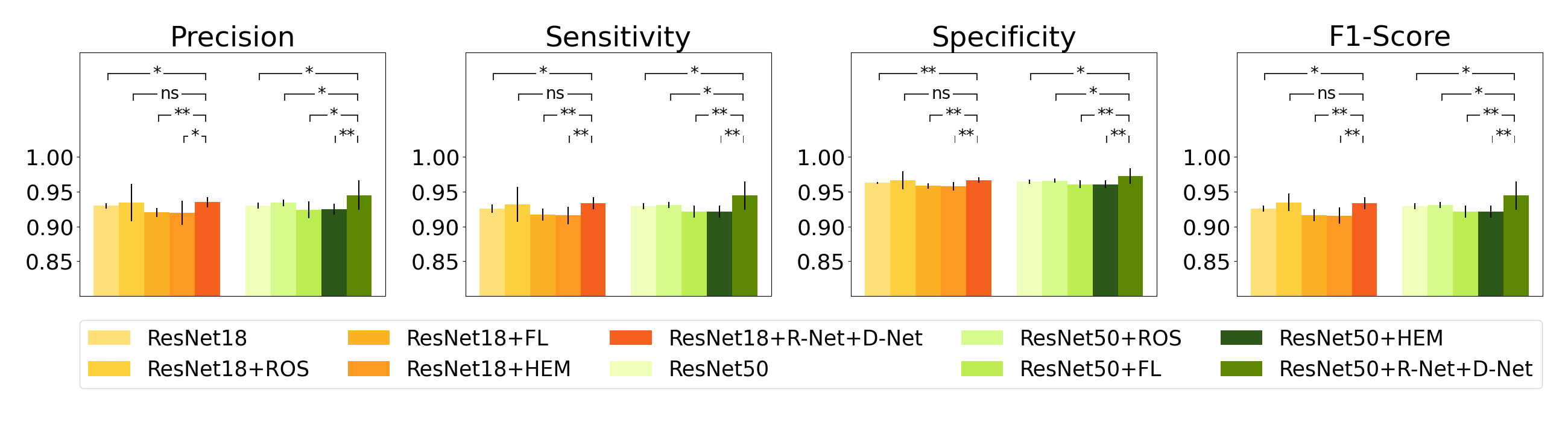}
    \caption{Classification performance of the proposed method and other methods for comparsion on the COVIDx dataset. 
    \textbf{ResNet18 or ResNet50}: vanilla ResNet used as F-Net;
    \textbf{+ROS}: train ResNet with ROS strategy; \textbf{+FL}: trained ResNet using focal loss; \textbf{+HEM}: trained ResNet with HEM strategy; \textbf{+R-Net+D-Net}: trained ResNet with our proposed method. The image size of the training data was $224\times 224$ pixels.
    Statistical significance symbols above the bars indicate the $\it{t}$-test results for the null hypothesis that the average performances of the selected methods are the same, where $ns$ represents $\it{p}$-value$>0.05$ (no significant difference); The symbols $*$ and $**$ indicates the $\it{p}$-value is less than $0.05$ and $0.01$ (with significant difference), respectively.
    } 
    \label{fig:NetEffect_COVID}
    \end{figure}

In the COVID-19 case study,
ResNet18 and ResNet50 were employed as the F-Net to evaluate the performance of the proposed method. The classification performance of using ResNet18 or ResNet50 without auxiliary networks was set as the baseline.
As shown in Figure~\ref{fig:NetEffect_COVID}, when using ResNet50 as the F-Net,
our proposed method demonstrates superior classification performance to three compared methods, ROS, FL, and HEM described in Section~\ref{subsec:comp}, in terms of precision, sensitivity, specificity and F1-score. 
Our method achieved 1.5\%, 1.5\%, 0.8\%, 1.5\% improvement in terms of precision, sensitivity, specificity, and F1-score, respectively, compared to the baseline.
An unpaired $\it{t}$-test was further performed between the proposed method and the baseline for assessing the stability of model training. 
The null hypothesis was defined as the average performance of the proposed method being equal to or lower than that of the baseline. 
The $\it{p}$-values for this test were $0.020$, $0.016$, $0.020$ and $0.016$ for the precision, sensitivity, specificity, and F1-score, respectively. 
The {\it{t}}-test comparison results show that the difference between our proposed method and baseline is significant, therefore, the performance improvement by using the adversarial learning strategy.
The proposed method shows high stability even when the dataset is class-imbalanced and contains visually-similar inter-class samples.
When the ResNet18 is employed as the F-Net,
the proposed method can still achieve significantly better performance compared to the other methods, except for the ROS method. 
However, the degree of performance improvement is smaller compared to that using ResNet50 as the F-Net, 
which implies that network depth affects the classification performance. 

\begin{figure}[h]
    \centering
    \includegraphics[width=\textwidth]{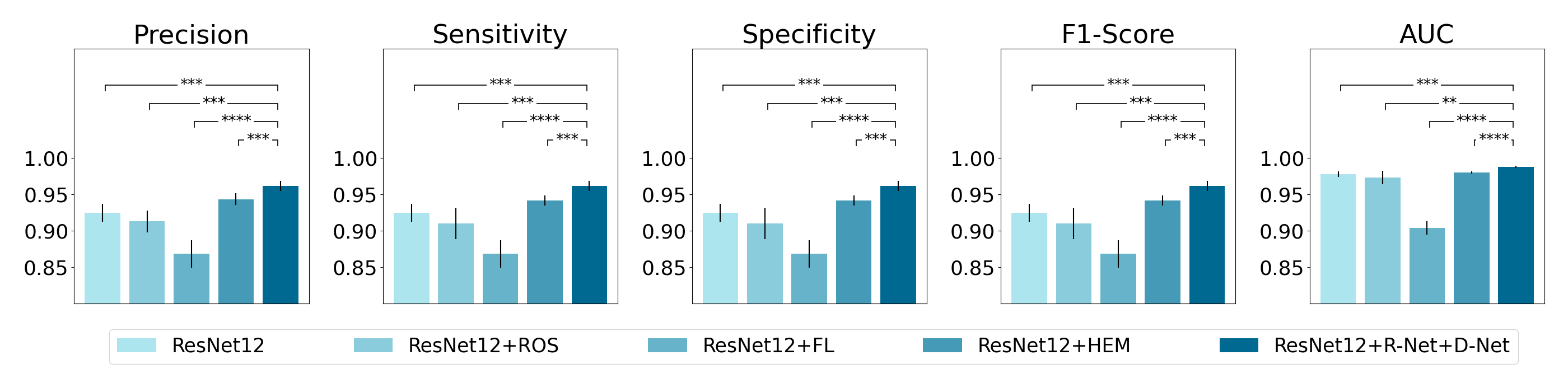}
    \caption{Classification performance of the proposed method and other methods for comparsion on the OPSCC dataset. 
    ResNet12 was used as the F-Net; Other symbols have the identical meaning as that in Figure~\ref{fig:NetEffect_COVID}
    The image size of the training data was $32\times 32$ pixels. The symbols $**$, $*{*}*$, and $*{**}*$ indicate the $\it{p}$-value is less than $0.01$, $0.001$, and $0.0001$, respectively
    } 
    \label{fig:NetEffect_HN}
\end{figure}

In the case study using the OPSCC dataset, the proposed method also achieves superior performance to all other methods for comparison. Similarly, the performance of the model using vanilla ResNet12 as the F-Net without AuxCNN was set as the baseline. As shown in Figure~\ref{fig:NetEffect_HN}, our method achieves a statistically significant 3.7\% improvement compared to the baseline in terms of precision, sensitivity, specificity, and F1-score. In addition, the AUC score is significantly improved from 97.94\% to 98.84\% by using our method.  
The results validate the effectiveness of our method in the binary classification task with small input image sizes. 

\subsection{Effects on classification performance of auxiliary networks}
\label{subsec:ablation}

Ablation experiments were performed to further quantify the benefit of incorporating the two proposed auxiliary networks in improving the classification performance. 
We compared the classification performances among models without AuxCNN, with R-Net only, and with both R-Net and D-Net simultaneously.

\begin{figure}[!h]
    \begin{subfigure}[b]{\textwidth}
        \centering
        \includegraphics[width=\textwidth]{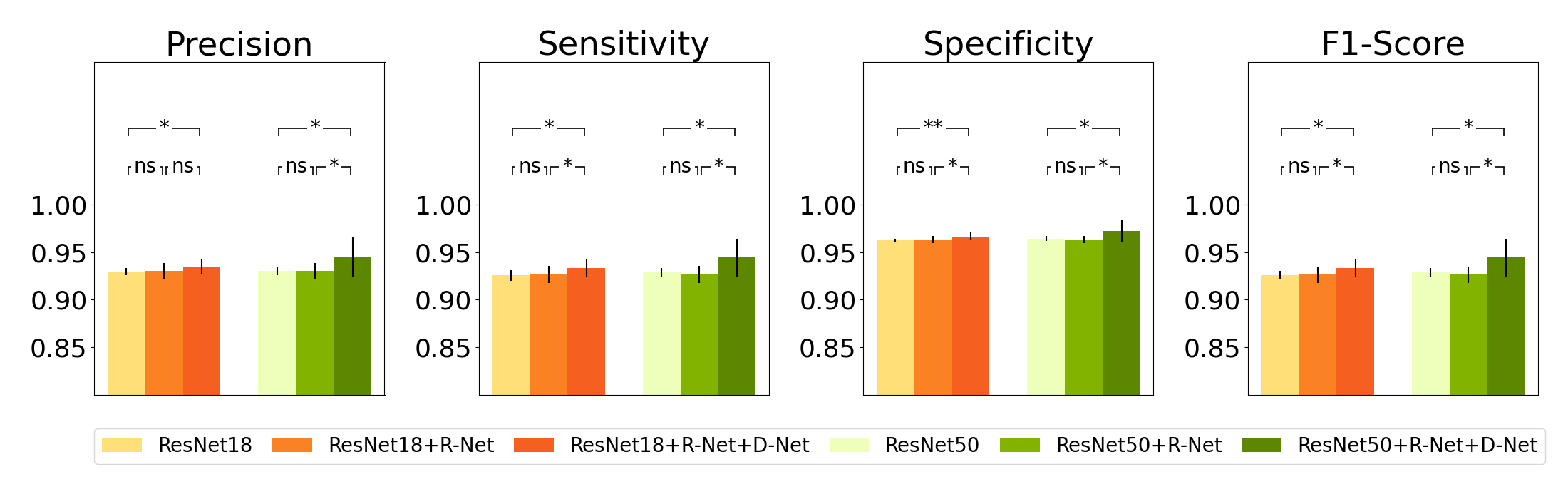}
        \caption{training image size: $224\times 224$}
        \label{fig:ablation_224}
    \end{subfigure}
    \begin{subfigure}[b]{\textwidth}
        \centering
        \includegraphics[width=\textwidth]{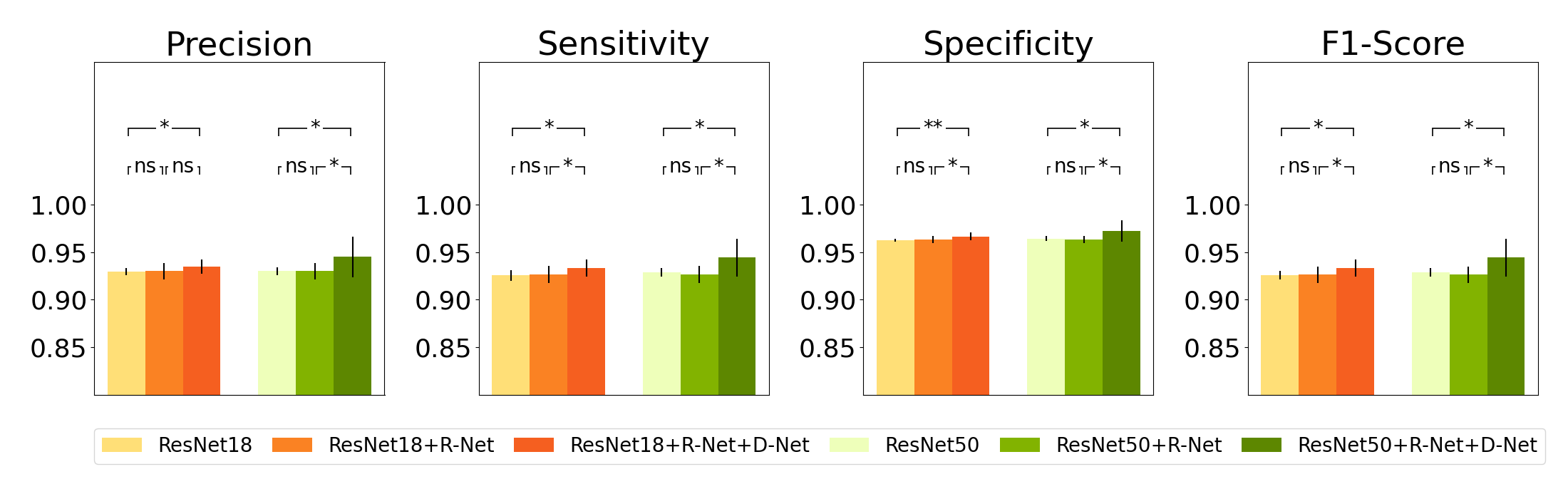}
        \caption{training image size: $448\times 448$} 
        \label{fig:ablation_448}
    \end{subfigure}
    \caption{Quantitative evaluation of using R-Net and D-Net in improving classification performance on the COVIDx dataset. The sizes of the training images are $224\times 224$ pixels in panel (a) and $448\times 448$ pixels in panel (b), respectively. 
    \textbf{ResNet18 or ResNet50}: the architecture of the ResNet as F-Net; 
    \textbf{+R-Net}: trained F-Net with the support of R-Net only; \textbf{+R-Net+D-Net}: trained F-Net with the support of both R-Net and D-Net. 
    \label{fig:ablation}
    }
    \end{figure}

Figure~\ref{fig:ablation_224} shows the contribution of each auxiliary network in the COVID-19 case study. 
When F-Net was set as ResNet50 and was trained with the support from both R-Net and D-Net, 
a significantly greater improvement in F1-score can be achieved ($1.5\%$ with $\it{p}$-value as $0.016$).
As expected, when the F-Net was trained with only support from the R-Net, no significant improvement was observed in terms of the metrics. 
This result indicates that D-Net plays an important role in optimizing the extracted feature of F-Net. 
A similar phenomenon was observed when the F-Net was set as ResNet18 but with a reduced degree of improvement.
However, as shown in Figure~\ref{fig:ablation_448}, when the training image size was up-scaled to $448\times 448$ pixels, 
the performance of F-Net trained with AuxCNN shows no significant improvement when using ResNet50 as F-Net, 
while a significant difference compared to the case using ResNet18 as the F-Net can still be observed. 
This result implies that both F-Net architecture and the size of training images can affect the effectiveness of the auxiliary networks, thus affecting the final classification performance.

\begin{figure}[!h]
    \centering
    \includegraphics[width=\textwidth]{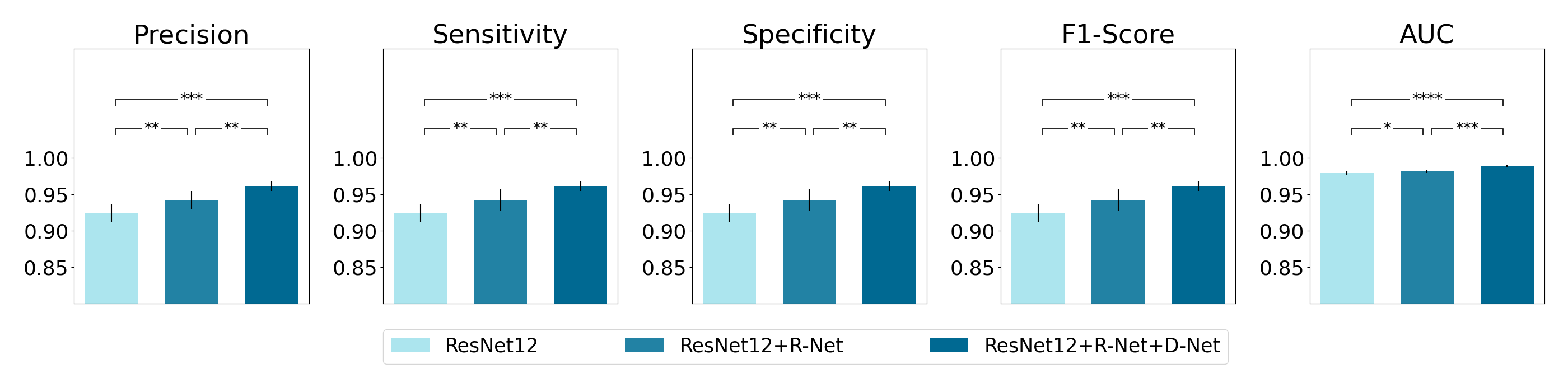}
    \caption{Quantitative evaluation of using R-Net and D-Net in improving classification performance on the OPSCC dataset. The training image size was $32\times 32$ pixels.
    ResNet12: the architecture of the ResNet as F-Net. 
    \textbf{ResNet12+R-Net}: trained F-Net with the support of R-Net only; \textbf{ResNet12+R-Net+D-Net}: trained F-Net with the support of both R-Net and D-Net. 
    \label{fig:ablation_hn}
    }
\end{figure}

The ablation experiment results in the OPSCC case study are shown in Figure~\ref{fig:ablation_hn}, both R-Net and D-Net could significantly improve the classification performance. The results demonstrate that the auxiliary networks perform better when the input image size is relatively small and the structure of the F-Net is relatively simple. 
It suggests that various factors such as F-Net architecture and the input image size may affect the effectiveness of each auxiliary network.

\subsection{Effect on performance of various network factors}
\label{subsec:factors}

Various factors that may affect the performance of the proposed network were investigated in the COVID-19 case study. They are discussed below. 

\subsubsection{F-Net network depth}
\label{subsubsec:network_depth}
\begin{figure}[h]
\centering
\includegraphics[width=\textwidth]{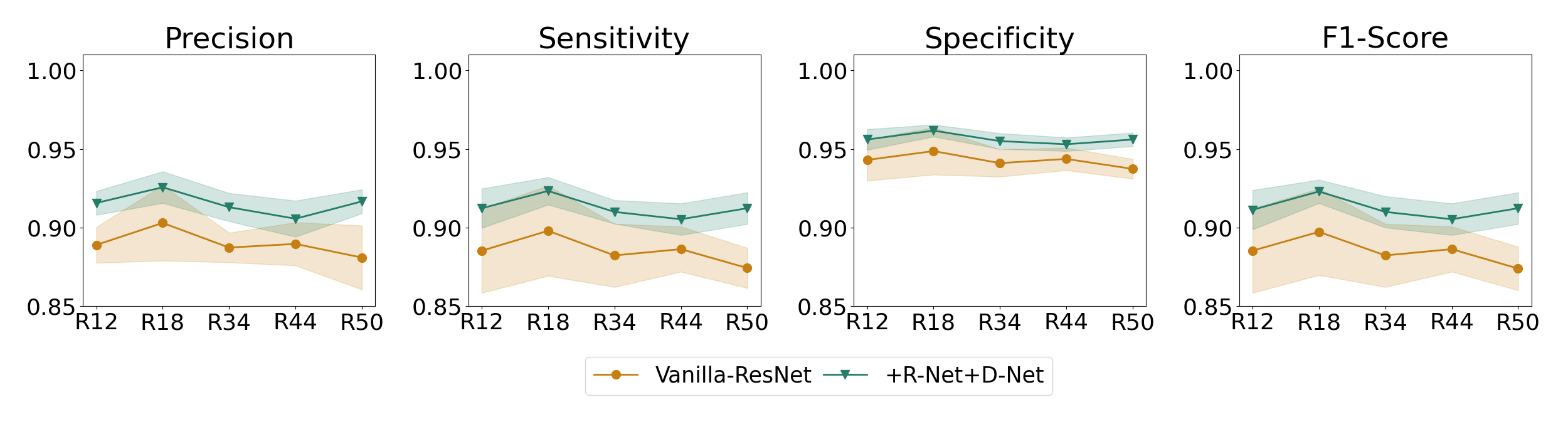}
\caption{
The effect of F-Net network depth on the classification performance. \textbf{Vanilla-ResNet}: the vanilla ResNet used as the F-Net, where R12, R18, R34, R44, and R50 represent ResNet12, ResNet18, ResNet34, ResNet44 and ResNet50, respectively. The larger the number is, the deeper the ResNet architecture is. \textbf{+R-Net+D-Net}: trained F-Net with our proposed method.
The shadow areas show the confidence interval bands of each trained F-Net. 
}
\label{fig:DepthEffectTFS}
\end{figure}

To investigate the variations of classification performance along with the network depths, a set of ResNets with different depths of \{12, 18, 34, 44, 50\} was employed as the F-Net. 
All the ResNets were trained with Xavier scheme~\cite{pmlr-v9-glorot10a} to initialize network parameters.
As shown in Figure~\ref{fig:DepthEffectTFS}, all of these F-Nets achieved significantly higher classification performance with our method compared to those based on vanilla ResNets, 
validating the versatility of the auxiliary learning strategy.
Notably, the ResNet18-based method achieves the best performance among all variations, 
while ResNet50 demonstrates the largest improvement by 3.6\%, 3.8\%, 1.9\%, 3.8\% in terms of all the evaluation metrics compared to its vanilla version.
This result indicates that a deeper network does not necessarily guarantee a better classification performance, 
and an appropriate network architecture should be determined according to the datasets and training strategy. 

\subsubsection{Effect of image size of training data}
\label{subsubsec:size}
\begin{figure}[!ht]
\centering
\includegraphics[width=0.75\textwidth]{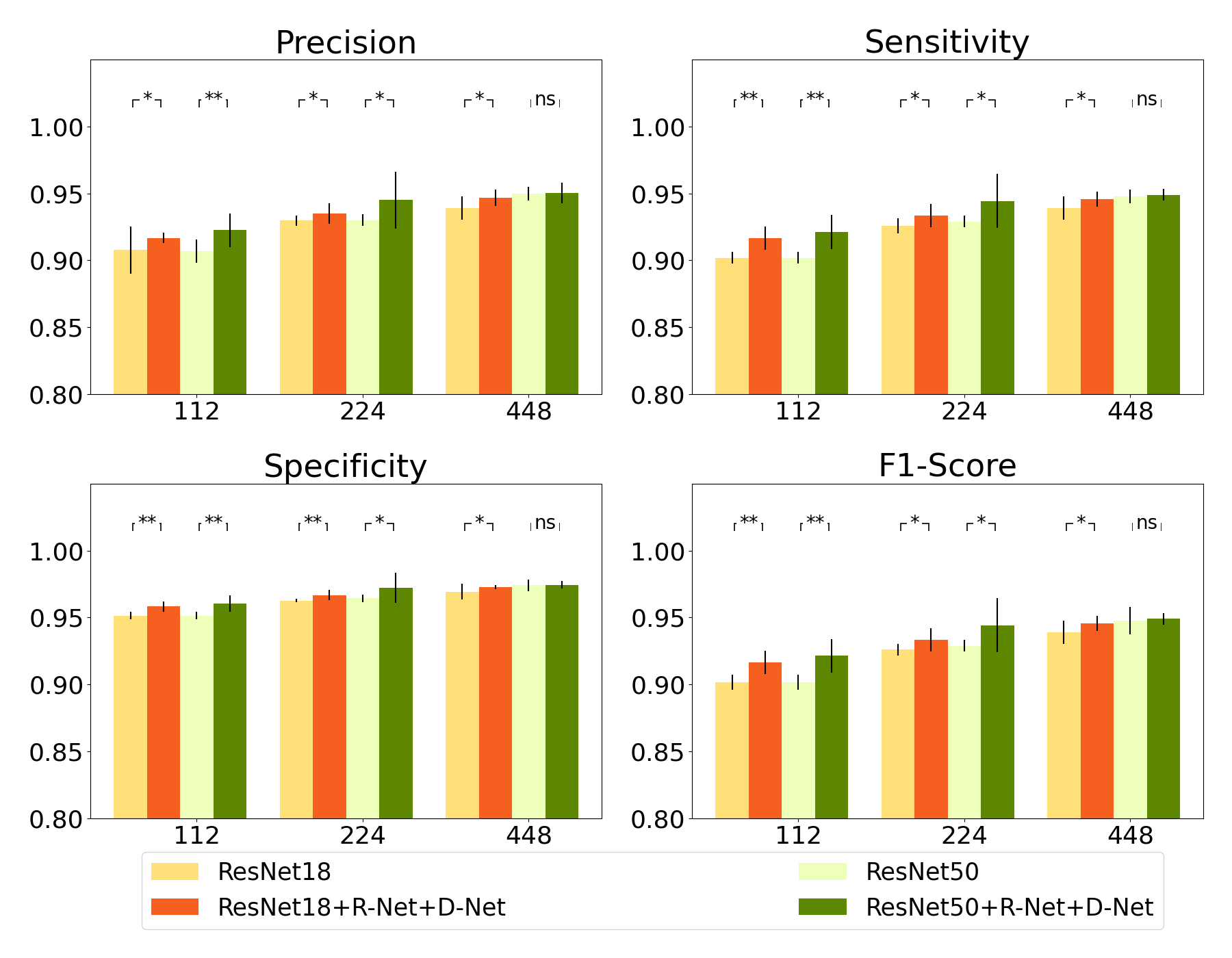}
\captionof{figure}{
The effect of the size of training images on the classification performance.
The number 112, 224, and 448 shown on the x-axis represent the input training image size was $112\times 112$ pixels, $224\times 224$ pixels, and $448\times 448$ pixels, respectively.}
\label{fig:SizeEffect}
\end{figure}

To investigate the effect of variations of training image size on model performance, 
the F-Net was trained using the images with three different sizes, $112\times 112$, $224\times 224$, and $448\times 448$ pixels as described in Section~\ref{subsec:net_hyper}.
Two F-Net architectures, ResNet18 and ResNet50, were used for evaluation, and the results are shown in Figure~\ref{fig:SizeEffect}.
When using ResNet18 as F-Net, our methods show universal effectiveness in improving the performance of F-Net across all input image sizes.
However, when using ResNet50 as F-Net, our method could achieve noticeably better classification performance 
on images with the size of $112\times 112$ or $224\times 224$, but trivial improvement for images with the size of $448\times 448$. 
These results indicate the effectiveness of our proposed method on various image sizes, though the degree of improvement may vary.  

\subsubsection{Training dataset size}
\label{subsubsec:data_size}

\begin{figure}[!h]
    \centering
    \includegraphics[width=\textwidth]{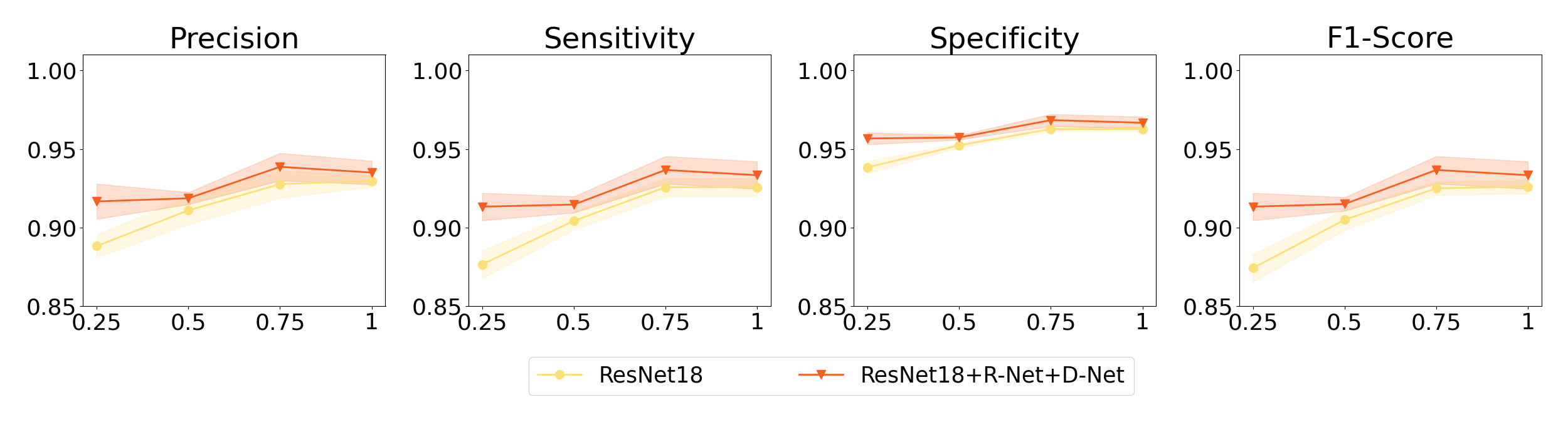}
    \caption{
    The effect of training dataset size on the classification performance. ResNet18 was used as the F-Net in this test. The numbers from 0.25 to 1 shown on the x-axis represent the ratio of the reduced number of training images. 
    }
    \label{fig:DataSizeEffect}
\end{figure}

Using a limited number of images for training would degrade the classification performance mainly due to the risk of overfitting~\cite{Alzubaidi2020TowardsAB}.
To understand how the model performance varies 
along with the number of training samples, 
25\%, 50\%, and 75\% of samples were randomly selected from each class of the original COVIDx dataset to re-train the networks, respectively.
As shown in Figure~\ref{fig:DataSizeEffect}, using a smaller training dataset could lead to performance degradation. 
However, our auxiliary learning strategy can alleviate the performance degradation compared to the models without the auxiliary networks when fewer training images are available. For example, when 25\% of the dataset was used for training, the vanilla ResNet18 trained without AuxCNN achieves 88.8\% of F1-score while it can achieve 93.0\% trained on the whole dataset, indicating a large performance degradation using a small dataset. 
Our method can reduce such performance degradation from 4.2\% to 2.1\%.
Interestingly, when the ratio of the training dataset reached 75\%, the network could achieve a similar performance as that of using the whole dataset.
These results indicate the effectiveness of the proposed method in improving the classification performance on a small dataset, 
though the degree of improvement varies with the dataset size.  

\subsubsection{Transfer learning (TL) method}
\label{subsubsec:tl}
\begin{figure}[!ht]
    \centering
    \includegraphics[width=\textwidth]{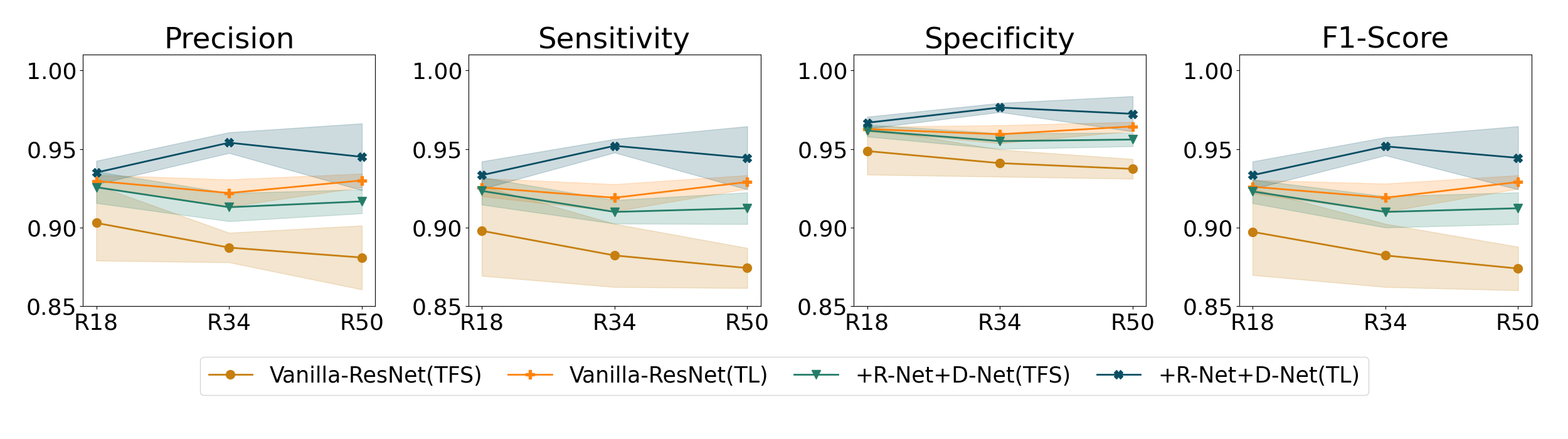}
    \caption{
    The effect of transfer learning strategy on classification performance. \textbf{TL}: with transfer learning; \textbf{TFS}: with training from scratch. Training image size was $224\times 224$. \textbf{Vanilla-ResNet}: the vanilla ResNet used as the F-Net, where R12, R34, and R50 represent the vanilla ResNet12, ResNet34, and ResNet50, respectively. \textbf{+R-Net+D-Net}: trained specific F-Net with our proposed method.
    }
    \label{fig:TLEffect}
\end{figure}

Transfer learning (TL) has been commonly employed to handle the issue of limited training dataset for training DL networks~\cite{shorten2019survey,Alzubaidi2021ReviewOD}.
The performances of three ResNet architectures, ResNet18, ResNet34, and ResNet50, were investigated with TL strategy or training from scratch (TFS).
The parameters of each network were initialized with the pre-trained weights on the ImageNet dataset~\cite{Deng2009ImageNetAL} in the TL strategy, 
or the Xavier initialization scheme~\cite{pmlr-v9-glorot10a} in the TFS strategy.
As shown in Figure~\ref{fig:TLEffect},
our proposed method universally improves the network performances, using either TL strategy or TFS strategy. 
The results also validate the effectiveness of TL strategy. 
Particularly, our proposed method could achieve greater performance improvement when using the TFS strategy. The reduced confidence band widths show improved training stability by using the auxiliary networks. 
Notably, similar model performances are observed by using either TL strategy or our method, showing a high potential to be employed if the pre-trained parameters are unavailable. 
This phenomenon implies that the auxiliary networks might work in a similar role to the TL strategy for regularization in improving training performance and stability.

\subsection{Visualization of reconstructed images of R-Net}
\label{subsec:visual}

\begin{figure}[!ht]
    \centering
    \includegraphics[width=\textwidth]{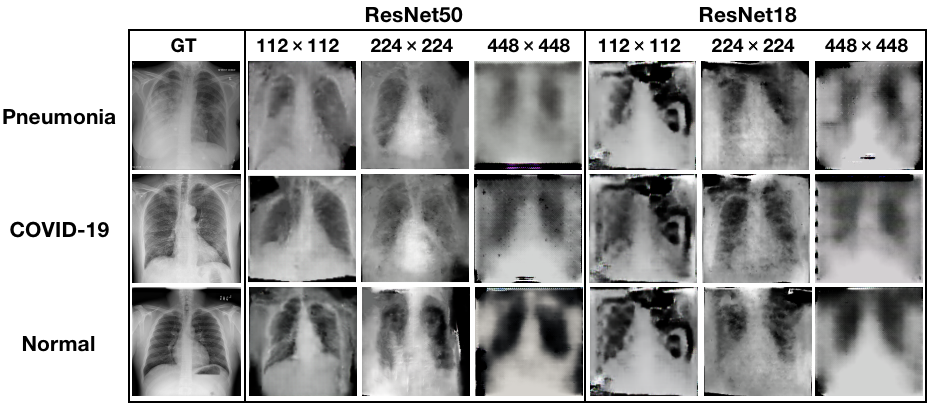}
    \caption{Visualization of the reconstructed images of COVID-19 examples. 
    The first column shows the ground truth images of each class label. 
    The second to the fourth columns show the reconstructed images using ResNet50 as the F-Net when the training image size is $112\times 112$, $224\times 224$, and $448\times 448$ pixels, respectively. 
    The last three columns show the reconstructed images when the F-Net is ResNet18.}
    \label{fig:covid-recon}
\end{figure}

We also visually evaluated the quality of R-Net reconstructed images considering the primary purpose of R-Net in the adversarial training process.
Hypothetically, the feature extracted by F-Net that achieves better classification performance should also restore the input image decently via R-Net. 
In the COVID-19 case study, however, as shown in Figure~\ref{fig:covid-recon}, 
the quality of reconstructed images from the R-Net roughly decreases as the image size increases, while the classification performance, on the contrary, increases as described in previous results.  
Particularly, when the F-Net was set as ResNet50 and the training image size was $112\times 112$, the reconstructed image showed the highest perceptual image quality.  
The larger the reconstructed image size was, the worse the reconstructed image quality was. 
Reduced image quality and severe artifacts were observed in the reconstructed images when using ResNet18 as the F-Net.

\begin{figure}[!ht]
    \centering
    \includegraphics[width=0.8\textwidth]{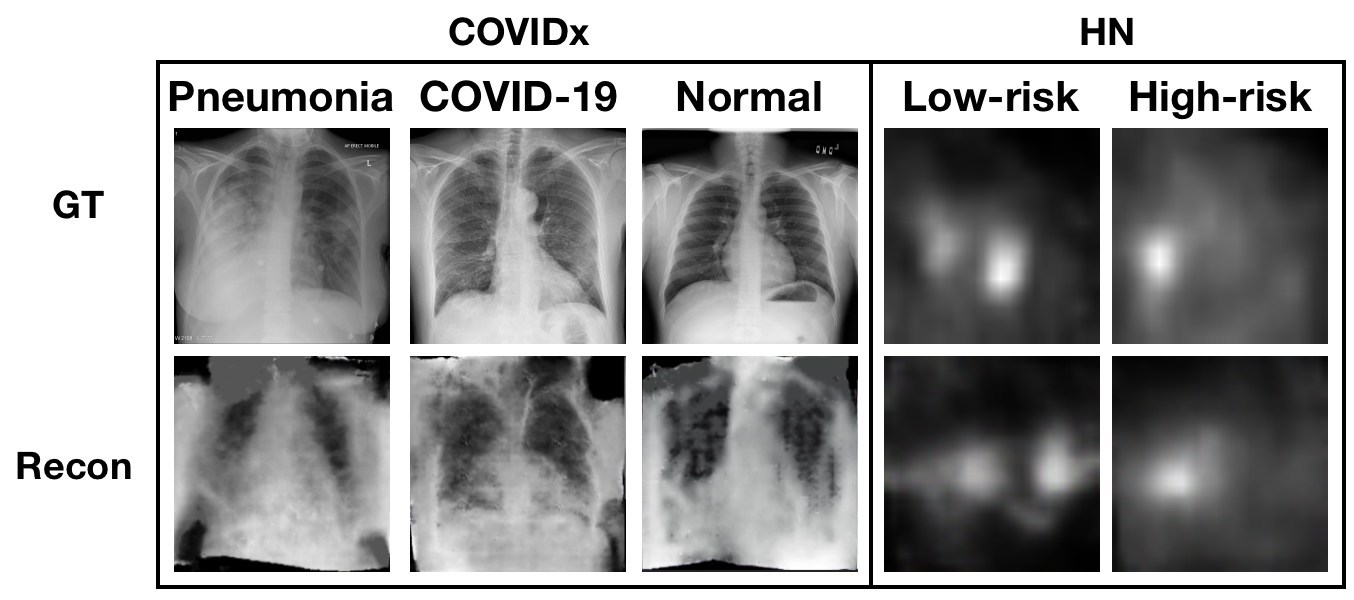}
    \caption{Visualization of the reconstructed images of COVID-19 and OPSCC examples using ResNet12 as F-Net. 
    The top row shows the ground truth images of each class label. 
    The bottom row shows the reconstructed images. 
    The image sizes of the COVID-19 examples were $224\times 224$ pixels, while the sizes of the OPSCC samples were $32\times 32$ pixels.}
    \label{fig:res12-recon}
\end{figure}

\begin{figure}[!ht]
    \centering
    \includegraphics[width=0.6\textwidth]{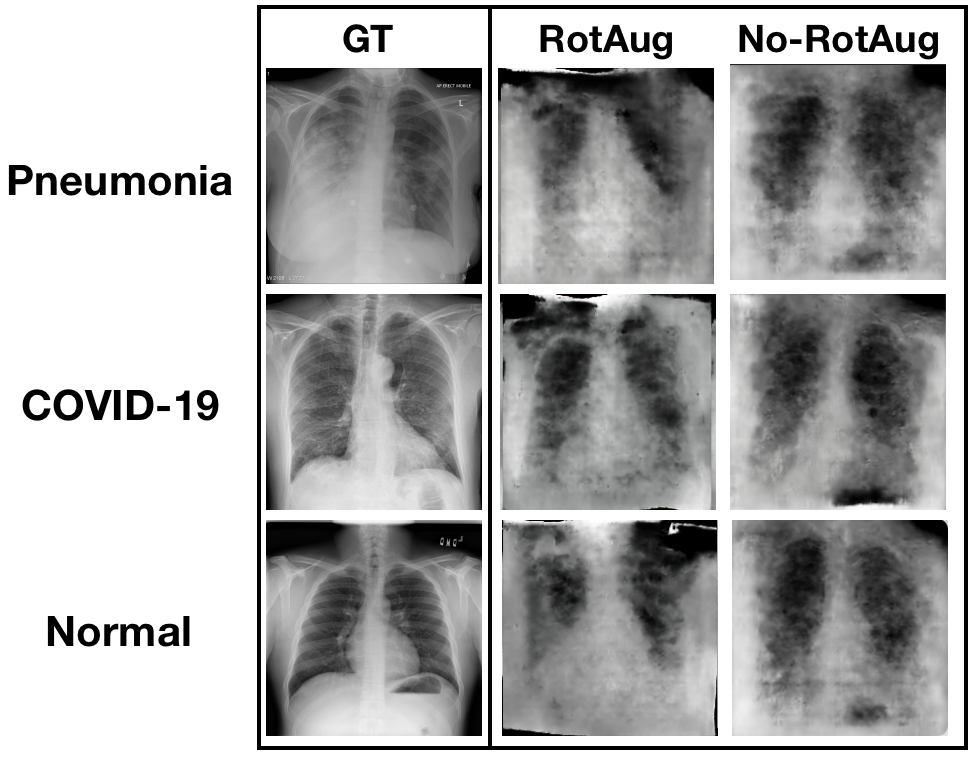}
    \caption{Visualization of the COVID-19 reconstructed images examples with or without random rotation for data augmentation during training. 
    \textbf{RotAug}: with random rotation augmentation; \textbf{No-RotAug}: without random rotation augmentation. The training F-Net was ResNet18 and training image size was $224\times 224$ pixels. }
    \label{fig:covid-recon-rot}
\end{figure}

These findings are counter-intuitive to our original assumption. 
One possible reason is that the extracted feature representation for better classification performance might be different from that for image reconstruction. 
Since the optimal model is obtained by assessing the classification performance of the model rather than reconstructed image quality,
the extracted feature may lose some semantic information for image reconstruction, although the classification performance is still improved.
Another possible reason is the limited decoding capability of the R-Net, which was originally proposed for low-resolution image (e.g. $64\times 64$ pixels) generation tasks.
This idea is supported by the reconstruction results in the OPSCC case study, as shown in Figure~\ref{fig:res12-recon}. 
No obvious artifacts are observed in the reconstructed images in the size of $32\times 32$ pixels.
Also, the random rotation applied in data augmentation might account for the artifacts around the image edges and corners. 
As shown in Figure~\ref{fig:covid-recon-rot}, the artifacts are reduced when this rotation trick was not introduced in the data augmentation procedure. 
This might be due to the extra zero-filling regions around corners after the original image was rotated, confusing the R-Net when approximating these regions. 
The interception of DL-based feature extraction (or called feature map) is an interesting research topic to gain a deep understanding of the deep-learning architectures.

\section{Discussion}
\label{sec:disc}

The performance of DL-based image classification methods can be affected by several data-related issues, 
such as limited training data, data imbalance, and extreme inter-class similarity.
To address these problems, a novel adversarial learning-based AuxCNN classification framework inspired by the adversarial concept of GAN is proposed in this study. 
There are two major innovations of the proposed framework.
First, the combination of GAN-based auxiliary networks and a hybrid loss-based adversarial training strategy enables the extraction of more salient features to improve the classification performance and alleviates overfitting by including not only inter-class discriminative information but also intra-class distribution information.
Second, the proposed modulized framework yields the flexibility of designing F-Net, R-Net, and D-Net with varied network architecture to achieve a better trade-off between the running speed and prediction accuracy under particular practical situations.

The usage of GAN for classification tasks has attracted lots of attention in the medical imaging analysis community~\cite{hu2018unsupervised,Varghese2017GenerativeAN,Waheed2020CovidGANDA, ma2020mri}.
Different from the other reported methods that disentangle the training of the GAN and the classifier,
we innovatively employed the GAN as an AuxCNN network to support the training to optimize the classification network. 
This technique facilitates the classification network to extract more informative features through the adversarial learning-based AuxCNN, even though the reconstructed image quality may not be optimal.

In addition, a hybrid loss strategy different from previous works where additional annotations are required~\cite{Ge2018ChestXC,Bentaieb2016MultilossCN,Li2020RobustSO,li2022task} was employed in this study.
For example, Aicha \etal employed a multi-loss strategy to combine segmentation loss and classification loss by expanding the classification network into a U-Net-like network~\cite{Bentaieb2016MultilossCN}.
By incorporating the semantic information of the object during classification, 
their method achieved good performance on a gland classification task using microscopy images.
Nevertheless, the hybrid loss strategy used in our study adopted the additional learning objectives from the image reconstruction process through adversarial learning, 
without the need for extra annotations in addition to the class labels in previous methods.

The proposed strategy of using adversarial learning-based AuxCNN and hybrid loss design provides a new application direction in classification, which also has great potential for other medical applications.
In the future, we will investigate the adaption and performance of the proposed framework for other medical imaging tasks such as lesion detection and segmentation~\cite{ma2020mri, geng20213d}, as well as under the circumstances where multi-modal medical image data is available~\cite{granstedt2022learned}. 
Besides, we will perform further studies to gain a deeper understanding of the mechanism of the proposed framework for given tasks.
In this study, various factors, including F-Net network depth, training image size, transfer learning strategy, and training dataset size, are investigated.
More potential factors, such as different network choices and loss functions, could be taken into consideration in future studies. 
For example, as an alternative to the ResNet, more advanced networks like EfficientNet~\cite{tan2019efficientnet}, CoAtNet~\cite{dai2021coatnet}, and Vision Transformer (ViT) ~\cite{dosovitskiy2020image} can be employed as the F-Net.
In addition, exploring the relationship between the reconstructed image quality and the classification performance can help reveal the mechanism of the proposed method. 
Larger datasets and more complex R-Nets, such as the generative network of Progressive GAN~\cite{karras2017progressive} or StyleGAN~\cite{Karras2019ASG}, might be employed to investigate the cause of reconstruction failure in the future study. 

\section{Conclusions}
\label{sec:conclusion}
A novel AuxCNN-based classification framework is proposed in this study. 
The GAN architecture and adversarial learning strategy are employed in the design of AuxCNN.
The proposed adversarial learning-based auxiliary networks and hybrid loss strategy enable the extracted feature to capture more representative and robust information for improving classification performance and stability. 
A three-class classification study on a COVID-19 dataset and a binary classification study on an OPSCC dataset were conducted to demonstrate the effectiveness of our proposed framework.
This modularized framework demonstrates its flexibility to adapt the feature extraction networks and auxiliary networks to various applications, thus holding the potential to be applied to other clinical applications in the future.

\section{Acknowledgment}

This work is original and has not been submitted for publication or presentation elsewhere. This work was
supported in part by NIH awards R01CA233873, R21CA223799,
Cancer Center at Illinois seed grant 2021, DoD Award No. E01 W81XWH-21-1-0062, and Health Care Engineering Systems Center, University of Illinois Jump ARCHES Award 2022. Conflict of interest: none declared.

\bibliographystyle{ieeetr}
\bibliography{manucript}

\end{document}